\documentclass[aps,pra,reprint,superscriptaddress,showpacs]{revtex4-1} 
\usepackage{graphicx}  
\usepackage{dcolumn}   
\usepackage{bm}        
\usepackage{verbatim}   
\usepackage[version=3]{mhchem} 
\usepackage{amsmath,amssymb}
\usepackage{textcomp} 
\usepackage{enumitem}
\usepackage{color}
\usepackage{psfrag}
\usepackage[breaklinks,colorlinks,citecolor=blue,linkcolor=blue,urlcolor=blue]{hyperref}
\usepackage{subfigure}

\usepackage{amsfonts}

\usepackage[raggedright]{titlesec}
\usepackage{titlesec}

\titleformat{\paragraph}
{\normalfont\normalsize\bfseries}{\theparagraph}{1em}{}
\titlespacing*{\paragraph}
{0pt}{3.25ex plus 1ex minus .2ex}{1.5ex plus .2ex}

\begin{document}

\title{Minimization of ion micromotion with artificial neural network}

\author{Yang Liu}
\thanks{These three authors contribute equally}
\affiliation{School of Physics and Astronomy, Sun Yat-Sen University, Zhuhai, 519082, China}
\affiliation{Center of Quantum Information Technology, Shenzhen Research Institute of Sun Yat-sen University, Nanshan Shenzhen 518087, China}

\author{Qi-feng Lao}
\thanks{These three authors contribute equally}
\affiliation{School of Physics and Astronomy, Sun Yat-Sen University, Zhuhai, 519082, China}
\author{Peng-fei Lu}
\thanks{These three authors contribute equally}
\affiliation{School of Physics and Astronomy, Sun Yat-Sen University, Zhuhai, 519082, China}
\author{Xin-xin Rao}
\affiliation{School of Physics and Astronomy, Sun Yat-Sen University, Zhuhai, 519082, China}
\author{Hao Wu}
\affiliation{School of Physics and Astronomy, Sun Yat-Sen University, Zhuhai, 519082, China}
\author{Teng Liu}
\affiliation{School of Physics and Astronomy, Sun Yat-Sen University, Zhuhai, 519082, China}
\author{Kun-xu Wang}
\affiliation{School of Physics and Astronomy, Sun Yat-Sen University, Zhuhai, 519082, China}
\author{Zhao Wang}
\affiliation{School of Physics and Astronomy, Sun Yat-Sen University, Zhuhai, 519082, China}
\author{Ming-shen Li}
\affiliation{School of Physics and Astronomy, Sun Yat-Sen University, Zhuhai, 519082, China}

\author{Feng Zhu}
\email[]{zhufeng25@mail.sysu.edu.cn}
\affiliation{School of Physics and Astronomy, Sun Yat-Sen University, Zhuhai, 519082, China}
\affiliation{Center of Quantum Information Technology, Shenzhen Research Institute of Sun Yat-sen University, Nanshan Shenzhen 518087, China}

\author{Le Luo}
\email[]{luole5@mail.sysu.edu.cn}
\affiliation{School of Physics and Astronomy, Sun Yat-Sen University, Zhuhai, 519082, China}
\affiliation{Center of Quantum Information Technology, Shenzhen Research Institute of Sun Yat-sen University, Nanshan Shenzhen 518087, China}

\date{\today}

\begin{abstract}

Minimizing the micromotion of the single trapped ion in a linear Paul trap is a tedious and timing-consuming work, but is of great importance in cooling the ion into the motional ground state as well as maintaining long coherence time, which is crucial for quantum information processing and quantum computation. Here we demonstrate that systematic machine learning based on artificial neural networks can quickly and efficiently find optimal voltage settings for the electrodes using rf-photon correlation technique, consequently minimizing the micromotion to the minimum. Our approach achieves a very high level of control for the ion micromotion, and can be extended to other configurations of Paul trap.

\end{abstract}

\maketitle

\vspace{-1mm}

\section*{Introduction}
Recently the wave of artificial intelligence has made great changes to industry, marketing and social life such as facial/voice recognition, prediction of consumer behaviors. Due to its great success, machine learning algorithms have been rapidly applied to diverse research subjects \cite{awad2015efficient}, including physics \cite{arsenault2014machine,rupp2015machine,li2015molecular,pilania2016machine,carrasquilla2017machine,raissi2018hidden,radovic2018machine,carleo2019machine,sarma2019machine,schutt2020machine}, chemistry \cite{goh2017deep,kitchin2018machine,butler2018machine,tkatchenko2020machine,cartwright2020machine}, biology \cite{tarca2007machine,angermueller2016deep,ching2018opportunities}, astronomy \cite{ball2010data,way2012advances,vanderplas2012introduction,ivezic2014statistics,baron2019machine}, ecnomics \cite{herbrich1999neural,mullainathan2017machine,athey2018impact} and finance \cite{bose2001business, khandani2010consumer, heaton2017deep, klaas2019machine}, weather forecast \cite{lai2004intelligent,sharma2011predicting,kendon2014heavier,choi2016prediction,holmstrom2016machine}, medical science \cite{kononenko2001machine}, material science \cite{raccuglia2016machine,ward2016general,ramprasad2017machine,sanchez2018inverse,schmidt2019recent} and drug discovery \cite{deo2015machine,obermeyer2016predicting,ching2018opportunities}. In physics, it is successful in solving problems ranging from high-energy \cite{radovic2018machine} and string theory to condensed matter \cite{arsenault2014machine,pilania2016machine,carrasquilla2017machine}, to astrophysics \cite{vanderplas2012introduction}, to atomic, molecular and optical physics \cite{seif2018machine,sarma2019machine}. Always, it play a significant role in reducing the complexity of the problem and in optimization of one or multiple experimental parameter with various mathematical algorithms. Especially, it can reduce a N-dimensional problem to only a few dimensions, which greatly simplifies the experimental realization of any NP-hard problem in physics, such as improve the fidelity of a quantum gate operation using trapped ions \cite{} for quantum computation and quantum information processing. However, it remains an uncharted territory and a mistery to physicists that, to what extend can it assist and boost the research. 

In current research, we try to minimize the micromotion of the trapped ion. The excess micromotion of a trapped ion can come from stray electric fields, RF pickup by one of the DC electrodes, and a phase difference between the RF electrodes. It causes the ion to oscillate at the RF frequency, resulting in Doppler shifts as the ion moves towards or away from the laser beam. This will reduce the efficiency of both cooling and state detection, and could heavily reduce the ion's lifetime. More generally, it deteriorate the harmonic character of the ion motion, making it impossible to construct a ion quantum gate with high fidelity. Therefore, it is important to eliminate the micromotion to the highest degree possible. 

Multiple techniques are widely used for minimizing the excess micromotion of trapped ions, such as sideband \cite{chwalla2009precision,doret2012controlling}, parametric excitation \cite{narayanan2011electric,tanaka2012micromotion}, photon-correlation \cite{1998JAP, 2015JAP} etc. All of them are quite successfully, photon-correlation especially, in terms of reach high-level of minimization of excess micromotion. However, most of these methods requires adjusting DC voltages repeatedly and is quite time-consuming, usually take hours and even days. Since there is no direct mathematical relation between DC voltages and the excess micromotion index $\beta$, and the minimization of the micromotion is clearly a multi-dimensional problem, which is a natural playground where machine learning plays a major role. Here we apply neural-network-based machine learning to fast solve the problem using photon-correlation technique.

\section*{Theoretical Scheme}
\paragraph{Excess Micromotion}

The trapping potential of an ion trap can be written as \cite{2003RvMP}
\begin{equation} \label{potential}
\Phi = \frac{V_{DC}}{R^2}(\alpha x^{2} + \beta y^{2} + \gamma z^{2}) + \frac{V_{RF}cos(\Omega t)}{R^2}(\alpha^{'} x^{2} + \beta^{'}  y^{2} + \gamma^{'}  z^{2}) 
\end{equation}
where the first term represents the dynamics confinement in Y-Z radial plane, and the second term represents the static confinement along the X axial direction. $V_{RF}$ and $\Omega$ is the amplitude and frequency of the RF oscillating electric field, respectively. $V_{DC}$ is the amplitude of the DC electric field, and $R$ is the distance between the potential center and the electrodes in the radial plane. We assume that the principle axes of the radial trapping potential are aligned along the directions $u_1$ and $u_2$, as shown in Figure ~\ref{electrodes}, and assume the equations of motion along each direction are fully decoupled, we can write the ion's equation of motion in the form of Mathieu equation \cite{2015JAP}
\begin{equation} \label{motion}
\ddot{u_{i}}+(a_{i}+2q_{i}cos(\Omega t))\frac{\Omega^{2}}{4}u_{i}=\frac{Q E_{dc, i}}{m}.
\end{equation}
Here $E_{dc, i}$ is the stray electric field along $u_i$ that effectively push the ion away from the RF null point, causing the RF null and the DC null to be in different positions. $Q$ and $m$ are the charge and mass of the ion, and the $q_i$ and $a_i$ are related to applied voltages. The solution to Eq.\ref{motion} to lowest order in $q_i$ and $a_i$ is 
\begin{equation} \label{solution}
u_{i}(t) \cong (u_{0,i}+u_{1,i}cos(\omega_{i}t + \phi_{si}))(1+\frac{q_{i}}{2}cos(\Omega t + \phi_{i})),
\end{equation}
where $u_{0, i}\cong \frac{4Q\bm{E}_{dc, i}}{m(a_{i}+\frac{q_{i}^{2}}{2})\Omega^{2}} \cong \frac{Q\bm{E}_{dc, i}}{m\omega_{i}^2}$ is a displacement of the equilibrium position from the rf node due to any residual static electric field, $u_{1, i}$ is the secular motion amplitude with the frequency of $\omega_{i} = \frac{\Omega}{2}\sqrt{a_{i}+\frac{q_{i}^{2}}{2}}$. $\phi_{si}$ is the phase of secular motion determined by the initial conditions of the ion position and velocity. $\phi_{i}$ is the phase of the micromotion. The motion corresponding to the $cos(\Omega t)$ term is driven by the applied ac electric field, and is called micromotion. From Eq.\ref{solution}, this micromotion consists of an intrinsic contribution from the periodic displacement due to the secular motion, and an excess micromotion with an amplitude $\frac{u_{0, i}q_{i}}{2}$ caused by the ac electric field at position $u_{0, i}$. 

With the assumption of $|u_{0, i}|\gg |u_{1, i}|$ and $|R\alpha \phi_{ac}|\gg |u_{1, i}|$, the amplitude of this excess micromotion \emph{$u^{'}$} can be written as \cite{1998JAP}
\begin{equation}
\mathbf{k} \cdot \mathbf{u}^{\prime}(t)=\beta \cos (\Omega t+\delta)
\end{equation}
where 
\begin{equation} \label{index}
\beta = \sqrt{(\frac{1}{2}\sum_{i=X^{\prime},Y^{\prime}}k_{i}u_{0, i}q_{i})^{2} + (\frac{1}{4}k_{X^{\prime}}q_{X^{\prime}}R\alpha \phi_{ac})^2}, 
\end{equation}
The excess micromotion can induce the first- and second-order Doppler effect, which will significantly alter the excitation spectrum of an atomic transition. Assuming the electric field of the excitation laser has the amplitude \emph{$E_0$}, frequency $\omega$, phase $\phi_L$, and wave vector \emph{$k$}, then this laser field can be expressed in the rest frame of the ion undergoing excess micromotion:
\begin{center}
\begin{equation} \label{electric_field}
\begin{aligned}
\mathbf{E}(t) &= Re{\mathbf{E}_{0} e^{i \mathbf{k} \cdot \mathbf{u} - i\omega_{L}t + \phi_{L}}}\\
		& \cong  Re{\mathbf{E}_{0} e^{i \mathbf{k} \cdot \mathbf{u_{0}+u^{\prime}} - i\omega_{L}t + \phi_{L}}}\\
		& \cong  Re \left\{\mathbf{E}_{0} e^{i \mathbf{k} \cdot \mathbf{u_{0}}} \sum_{n=-\infty}^{\infty}J_{n}(\beta)\right.\\
		& \left.\ \times e^{- i\omega_{L}t + \phi_{L} + i\cdot n\cdot (\Omega t + \delta + \frac{\pi}{2})}\right\},
\end{aligned}
\end{equation}
\end{center}
Applying Fourier transform to Eq.(\ref{electric_field}), we get \cite{riehle2006frequency}
\begin{equation} \label{Fourier}
\begin{aligned}
E(\omega) \propto J_{0}(\beta)\delta (\omega - \omega_{L}) + J_{1}\left({\delta (\omega - \omega_{L} + \Omega)e^{-i(\delta + \frac{\pi}{2})}} \right.\\
\left. {\- \delta (\omega - \omega_{L} - \Omega)e^{i(\delta + \frac{\pi}{2})}}\right) \\
\end{aligned}
\end{equation}
Consequently, the detected fluorescence can be written as \cite{2015JAP}
\begin{equation} \label{Fitting}
\begin{aligned}
&S(\Delta, t)=\left|\int_{-\infty}^{\infty} A\left(\omega-\omega_{0}\right) E(\omega) e^{i \omega t} d \omega\right|^{2} \\
& \propto J_{0}^{2}(\beta)|A(\Delta)|^{2}\\
& +J_{1}^{2}(\beta)\left(\left|A\left(\Delta+\Omega_{rf}\right)\right|^{2}+\left|A\left(\Delta-\Omega_{rf}\right)\right|^{2}\right) \\
& +2 J_{0}(\beta) J_{1}(\beta)\left|A^{*}(\Delta) A\left(\Delta+\Omega_{rf}\right) - \right.\\
& \left.\ A(\Delta) A^{*}\left(\Delta-\Omega_{rf}\right)\right| \cos \left(\Omega_{rf} t+\varphi\right) \\
&+2 J_{1}^{2}(\beta)\left|A\left(\Delta+\Omega_{rf}\right) A^{*}\left(\Delta-\Omega_{rf}\right)\right| \cos \left(2 \Omega_{rf} t+\varphi^{\prime}\right),
\end{aligned}
\end{equation}
where $\Delta = \omega_{L} - \omega_{0}$ is the detuning of the excitation laser frequency from the atomic resonance, the phase $\varphi=\arg \left(A^{*}(\Delta) A\left(\Delta+\Omega_{\mathrm{rf}}\right)-A(\Delta) A^{*}\left(\Delta-\Omega_{\mathrm{rf}}\right)\right)$, and $\phi_{'} = arg(A(\Delta + \Omega)A^{*}(\Delta - \Omega))$. 

\paragraph{Neural-Network-based Machine Learning}

Machine learning algorithms are typically divided into three categories: (1) Supervised Algorithms which uses labeled datasets to train algorithm to construct a model, which can predict new data \cite{mohri2018foundations, russell2002artificial}. It includes neural network, support vector machine, k-nearest neighbors, decision Tree, random forest, and so on. (2) Unsupervised Algorithms which learn from unlabeled data \cite{hinton1999unsupervised}. It can divide data into different groups via learning. (3) Reinforcement Algorithms which is about how intelligent agents take action to maxmize the reward \cite{kaelbling1996reinforcement}. In this paper, we try to solve a object-oriented problem which is minimizing a physical parameter, the micromotion index of a trapped ion, as much as possible, thus we apply first category machine learning algorithm, neural-network-based machine learning to one of the most fundamental researches in quantum mechanics, the motional study of driven harmonic oscillator in mixed external fields.  

One of the most successful techniques to attack such a high-dimensional problem is the artificial neural networks (16). They can perform exceedingly well in a variety of scenarios ranging from image and speech recognition (17) to game playing (18). Very recently, applications of neural networks to the study of physical phenomena have been introduced (19–23). Although Applying neural network technique has been used in few cases, it is still of fundamental and practical interest to solve such a high-dimensional problem using neural network based machine learning method. This ability could then be used to minimize ion's excess micromotion for other types of Paul traps, and to solve other high-dimensional problems that is commonly timing-consuming and hard to address experimentally. 

Here we introduce an artificial neural network specified by a set of internal parameters \emph{${U_{i}}$}, which represent the input for the machine learning, i.e. the voltages on related compensation electrodes. We present a stochastic framework for reinforcement learning of the parameters \emph{${U_{i}}$}, allowing for reaching the optimal minimization of ion's excess micromotion. The parameters of the neural network are optimized (in other words, trained) by static variational Monte Carlo sampling (24). We validate the accuracy of this approach by measuring the magnitude of the residual excess micromotion right after we apply the learning results (i.e. the output voltages) on the corresponding electrodes. The power of the neural-network based machine learning is demonstrated, obtaining state-of-the-art level of minimization in excess micromotion \cite{1998JAP, 2015JAP}. 

Here we introduce a real-time progressive automatic algorithm to find out the null point of our blade trap. Previous to this work, we find a set of voltages $U_{i, c}$ through position compensation, which is ramping up the RF power and keep the ion's position by adjusting voltages when we ramp down the RF power. Now, we generate a Gaussian distribution of voltage for each compensation electrode using $U_{n, c}$ as the center and $\Delta U_{n}$ as the standard deviation. Thus we have a large sets of voltage combination. Then, we apply these voltage sets to the electrodes one at a time, and measure their corresponding photon-correlation spectrum in $x$ and $y$ direction. Following this, we obtain the micromotion index $\beta_x$ and $\beta_y$ for each voltage set by fitting each spectrum to the equation (~\ref{spectrum}). Based on these voltage sets and their indices, a model can be learned using neural network. Then it can give us a set of voltages $U_{n, m=1}$ which allow for the minimum $\beta_x$ and $\beta_y$ rather straightforward using anneal algorithm. Consider the above steps as a cycle/block, we repeat it a number of times, until we find out the actual null point. This is varified by the $\delta (\beta_{x}) < \epsilon $ and $\delta (\beta_{y}) < \epsilon $. 

In our neural network algorithm, we classify these artificial neurons into three class: $input$, $hidden$, $output$. Each artificial neuron has input and output. For $input$ $neurons$, input is dataset, output is the same of input. For the first $hidden$ $layer$ $neurons$, input is the output of $input$ $neurons$, output is data after linear transformation and nonlinear transformation. For the other $hidden$ $neurons$, input is the output of former $hidden$ $neurons$, output is also the result of transformation. For $output$ $neurons$, input is the same of output - the output of former $hidden$ $neurons$. 

\begin{figure}
	\centering
	\subfigure{\includegraphics[width=3.3in]{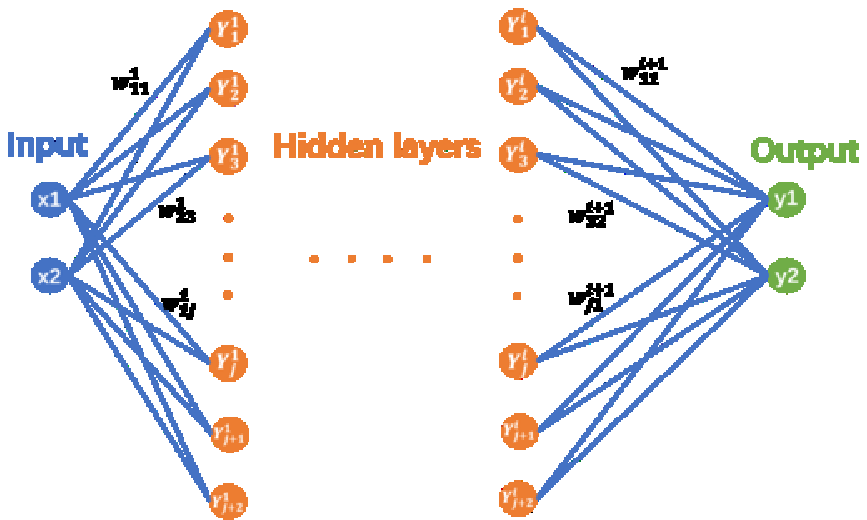}}\vspace{-10pt}
	\caption{Basic configuration of the neural network.}
	\label{NN}
\end{figure}

In Figure ~\ref{NN}, we show the basic configuration of an artificial neural network, which include input $x1$ and $x2$, and hidden layer $Y_{j}^{i}$, and output $y1$ and $y2$. The neuron in the hidden layer and the output can be expressed by:
\begin{equation}
\begin{aligned}
&Y_{j}^{1} = f_{j}^{1}(\sum_{n=1}w_{nj}^{1}x_{n})+b_{j}^{1}, \\
&Y_{j}^{i} = f_{j}^{i}(\sum_{n=1}w_{nj}^{i}Y_{n}^{i-1})+b_{j}^{i}, \\
&y_{j} = f_{j}(\sum_{n=1}w_{nj}^{i+1}Y_{n}^{i})+b_{j}, 
\end{aligned}
\end{equation} 
where $f(x)$ is activation function. In general, we use rectified linear unit (ReLU) function \cite{nair2010rectified}, that is $f(x) = max(0,x)$. In this way, we can write down the expression for output neuron with other neurons. After giving all the biases and weights, we can calculate value of all $output$ $neurons$. With input dataset $ (x_i; f_i)$, we now have the prediction $(x_i; y_i)$ by neural network and $(x_i; \tilde{y}_i)$ by measurement. Then we define loss function, also called cost function, or mean squared error (MSE) function
\begin{equation}
	L(y,p) = MSE(\tilde{y},y) =  \frac{1}{n}\sum(\tilde{y}_i - y_i)^2
\end{equation}

In order to obtain a suitable model to fit the data, we need the loss function as small as possible. Typically,  backpropagation (BP) algorithm \cite{goodfellow2016deep}, combined with stochastic gradient descent (SGD) algorithm \cite{bottou1998online, kiwiel2001convergence}, is widely used for training artificial neural networks, i.e. to iterate the neural network until the loss function small enough in neural network \cite{orr2003neural}. 

\begin{figure}
	\centering
	\subfigure{\includegraphics[width=3.3in]{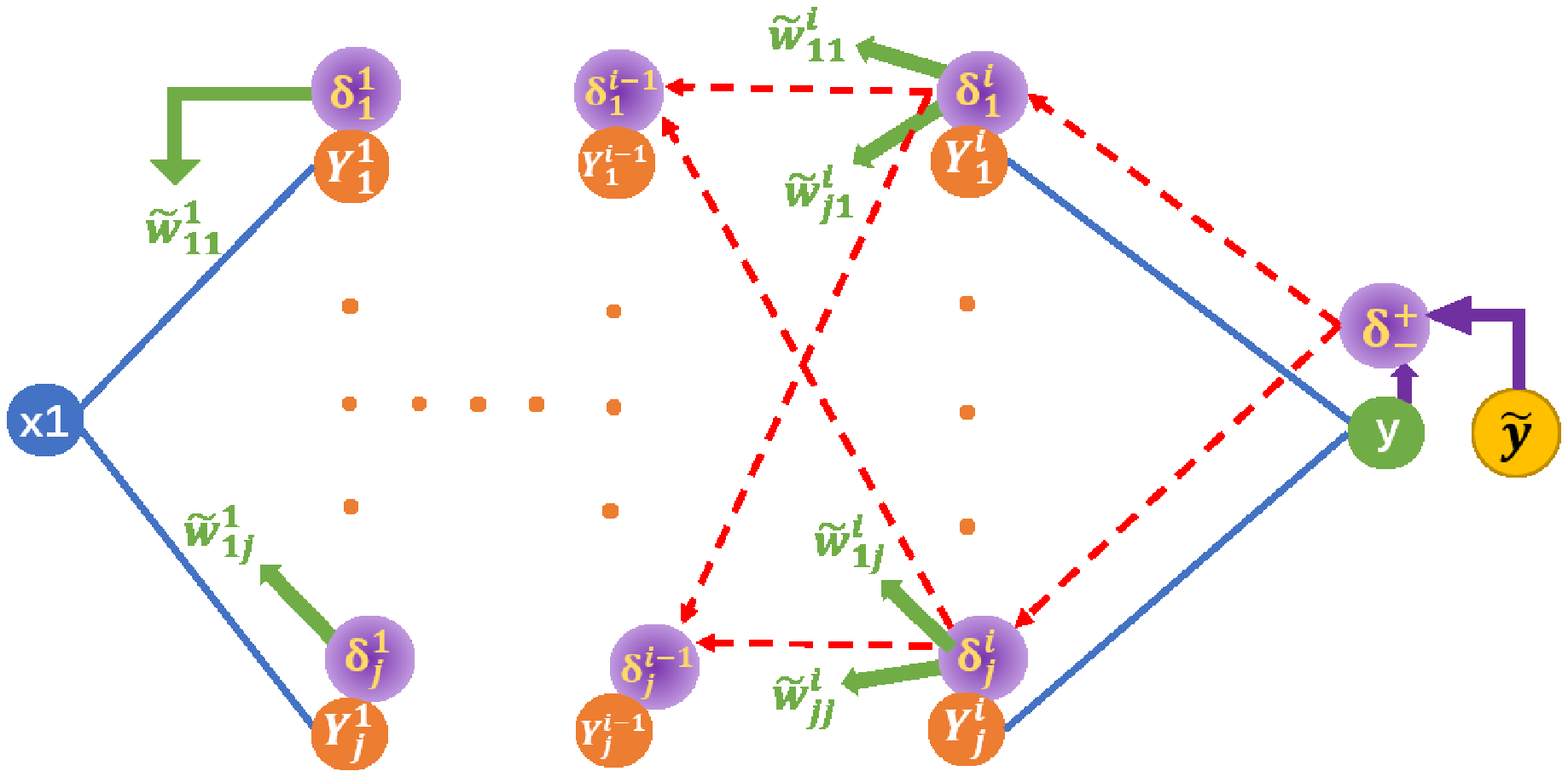}}\vspace{-10pt}
	\caption{Schemetics for the backpropagation algorithm.}
	\label{BP}
\end{figure}

Figure ~\ref{BP} shows the schematics for the backpropagation algorithm. First, we define the error $\delta = \tilde{y} - y$ as the difference between the deduced value from the measurement and the predicted value by the learning machine based on the constructed neural network. Assuming the hidden network has $i$ layer with $j$ neurons for each layer, the error in $y1$ would be propagated backwards according to
\begin{equation}
\left\{
             \begin{array}{lr}
             \delta_{n}^{i}(y1) = w_{n1}^{i+1}\dot \delta(y1), & 1\leq n\leq j. \\
             \delta_{n}^{k}(y1) = \sum_{m=1}^{j}w_{nm}^{k+1}\dot \delta_{m}^{k+1}(y1), & 1\leq k\leq i,1\leq n\leq j.\\
             \end{array}
\right.
\end{equation}
then the weight will be modified by 
\begin{equation}
\left\{
             \begin{array}{lr}
             \tilde{w}_{mn}^{k} = w_{mn}^{k} + \eta \frac{d L}{d w_{mn}^{k}}= w_{mn}^{k} + \eta \delta_{n}^{k} \frac{d f_{n}^{k}(x)}{d x} x_m, & k=1. \\
             \tilde{w}_{mn}^{k} =  w_{mn}^{k} + \eta \delta_{n}^{k} \frac{d f_{n}^{k}(x)}{d x} Y_{m}^{k-1}, & 2\leq k\leq i.\\
             \tilde{w}_{mn}^{k} =  w_{mn}^{k} + \eta \delta \frac{d f_{n}^{k}(x)}{d x} Y_{m}^{k-1}, & k= i+1.\\
             \end{array}
\right.
\end{equation}
So far, we have constructed a BP neural network, in practice, we just need to iterate the BP neural network until the loss function become small enough to get an appropriate model.

\section*{Experiment}
\begin{figure}[htbp]
	\centering
	\subfigure{\includegraphics[width=3.3in]{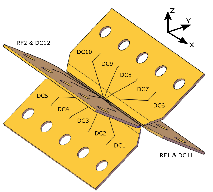}}\vspace{-10pt}
	\subfigure{\includegraphics[width=3.3in]{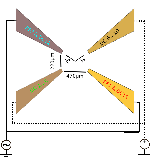}}\vspace{-10pt}
	\caption{(a) The five-segment blade trap used in our experiment. (b) The sketch-up diagram of the segmented blade trap. }
	\label{electrodes}
\end{figure}

In this section, we show how to implement the neural network to realize the minimization of the excess micromotion using the machine learning approach. First, we briefly introduce the experimental setup. The ion trap used in this experiment is similar to \cite{2016RSc}. It has four gold-plated ceramic blade electrodes, whose edges are parallel to the longitudinal (x) axis of the trap as shown in Figure ~\ref{electrodes}. Two opposite blades are driven with an RF potential with respect to the other two static blades, creating the transverse (y-z) quadrupole confinement potential. Appropriate static potentials provided by the longitudinal five-segment static blades serve to confine the ion along the x-axis. The RF electric quadrupole potential near the center of the trap is set by the RF amplitude on the trap electrode $V_0$, and can be closely approximated by $V_{RF}(y, z) = \frac{\kappa V_{0}}{2R^{2}}(y^{2}-z^{2})$. The gap between the electrodes is $470 \mu m$ and $220 \mu m$ in $y$ and $z$ direction, respectively. The distance from the trap center to the electrodes $R \approx 259 \mu m$. Other parameters in our blade trap are: $V_{RF1, RF2}\approx 250V$, $V_{DC1, DC2}\approx 11.75V$, $V_{DC3}\approx 1.742V$, $V_{DC4, DC5}\approx 9.35V$, $V_{DC6, DC7}\approx 12V$, $V_{DC8}\approx 1.9$, $V_{DC9, DC10}\approx 9.6V$, $V_{RF1}^{bias}\approx -0.079V$, $V_{RF2}^{bias}\approx -0.22V$, and $\Omega_{RF} = 22.625MHz$. With this configuration we achieved trap frequency of $\omega_{x},\omega_{y},\omega_{z} = 724kHz, 1.32MHz, 1.709MHz$ with 0.45W RF power. 

We measure the photon arrival time as a histogram using RF-photon correlation method, from which excess micromotion index $\beta$ can be obtained by fitting the histrogram profile to equation (1).  We supply the machine with the voltages on multiple DC compensation electrodes as the input and the index $\beta$ as the target, then we challenge the machine to discover the hidden high-dimensional voltage-to-$\beta$ surface. This relation between DC voltages and $\beta$ will give us a set of DC voltages, allowing optimal minimization of excess micromotion, which corresponds to the null point in the electric field. Since $beta_y$ and $beta_z$ in radial direction are strongly coupled together, the smaller $beta_y$, the smaller $beta_z$, only $beta_x$ in axial direction and $beta_y$ in radial direction instead of all three $beta$ in three Cartesian coordinates are measured. In the experiment, the detection laser is alternating in $x-$ and $y-$ direction for detection of $beta_x$ and $beta_y$ using photon-correlation technique. 

\begin{figure}
	\centering
	\subfigure{\includegraphics[width=3.3in]{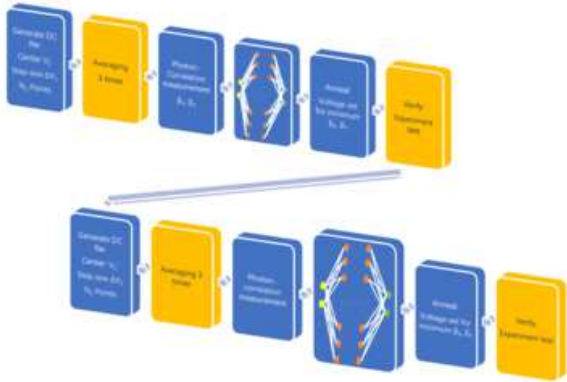}}\vspace{-10pt}
	\caption{One circulation loop in theoretical scheme for minimization of micromotion index ${\beta_x, \beta_y}$ using neural-network based machine learning.}
	\label{scheme}
\end{figure}

We take a two-input to two-output as an example, where two inputs are voltages on DC2 and DC7 respectively, and two outputs are $\beta_x$ and $\beta_y$ respectively. The procedure is carried out with multiple loops until predicted voltage set given by the neural network allow for the desired low level of micromotion. One of the loops is shown in Figure ~\ref{scheme}. It has two blocks (upper and lower panel) with four steps in each block. First, we generate a series of voltage set in a DC file, with $V_2$ and $V_7$ in the voltage sets having a Gaussian distribution in a large range. The center corresponds to the voltage set by the position compensation \cite{1998JAP}, which is monitoring the ion position $P(x_{0},y_{0},z_{0})$ with an EMCCD camera (iXon 897, Andor) at a high RF voltage, then lowering RF voltage and moving ion position to point $P$ by adjusting DC voltages. The standard deviation of the distribution are from a roughly estimatation of stable region based on Mathieu equation. After generating the DC file, we send the voltage sets to the electrodes in series through programmable precision DC voltage power supply (BS1-16-14, Stahl-electronics). Then we detect the fluorescence modulation in both x-direction and y-direction by alternately blocking the detection laser in y-direction and x-direction, using a START-STOP time-to-digital converter (HRM-TDC, SensL) to which photon numbers are fed from a photomultiplier tube (H10682-210, Hamamatsu). Each modulation spectra is taken three times to minimize the effects of drifting laser frequency and fluctuating laser power as well as the RF noise. 

After this, we deduce the $\beta_x$ and $\beta_y$ for each voltage set by fitting Eq.(\ref{Fitting}) to the measured spectra. Consequently, we use neural network to construct an appropriate model for minimization of excess micromotion. This is followed by the annealing process, which finds out the best voltage set allowing for minimum $\beta_x$ and $\beta_y$ simultaneously.  Since the range of the voltage sets in first block is relatively large, in order to have a precise voltage set and a better minimization, we repeat the process in the second block, using the voltage set attained from the first block as the center and typically half region of the first block to generate the new distribution of voltage set. 

\begin{figure}
	\centering
	\subfigure{\includegraphics[width=3.3in]{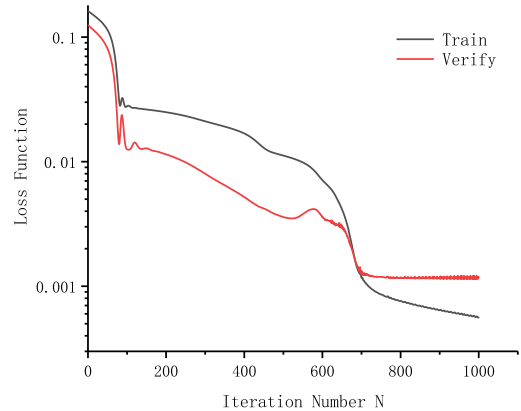}}\vspace{-10pt}
	\caption{The variation of loss function with iteration number N during the training of the learning model for $\beta_y$. Black curve represents the loss function between measurement and the prediction from the model, while the red curve represents the loss function between verification measurement and the prediction from the model. }
	\label{loss}
\end{figure}

For current learning model, we use 4 hidden layers, each having 100 neurons. The data flow is graphically represented in Figure ~\ref{scheme}. The output sequence of $\beta_{x}(V_{1},V_{2}...V_{i})$ and $\beta_{y}(V_{1},V_{2}...V_{i})$ from our model is then compared with the measured sequence $\tilde{\beta}_{x}(V_{1},V_{2}...V_{i})$ and $\tilde{\beta}_{y}(V_{1},V_{2}...V_{i})$ over the whole parameter space to evaluate the loss function. Once the loss function reach a value which is smaller than the threshold $\epsilon$, we conclude the model is appropriate. 

\begin{figure}
	\centering
	\subfigure{\includegraphics[width=3.3in]{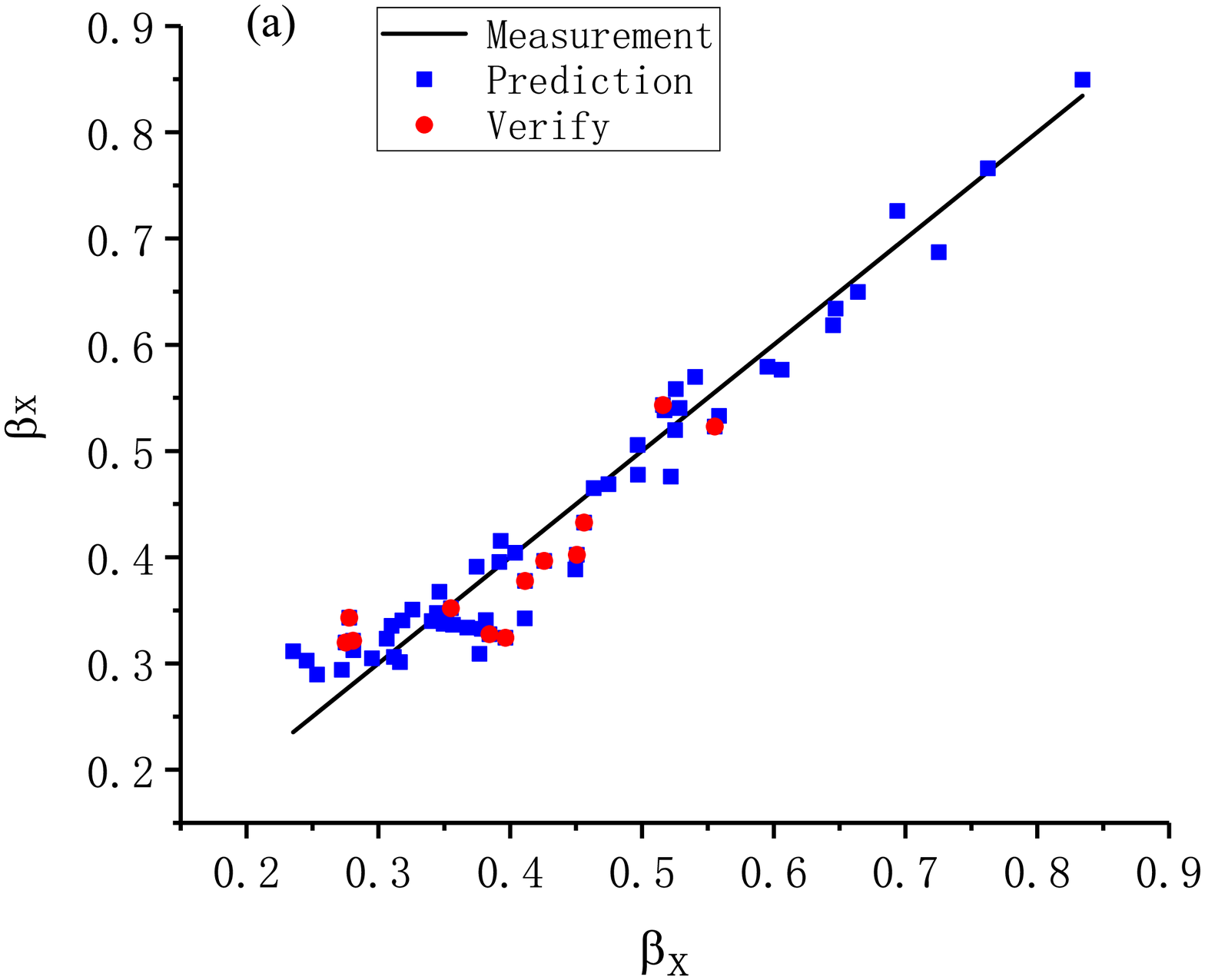}}\vspace{-10pt}
	\subfigure{\includegraphics[width=3.3in]{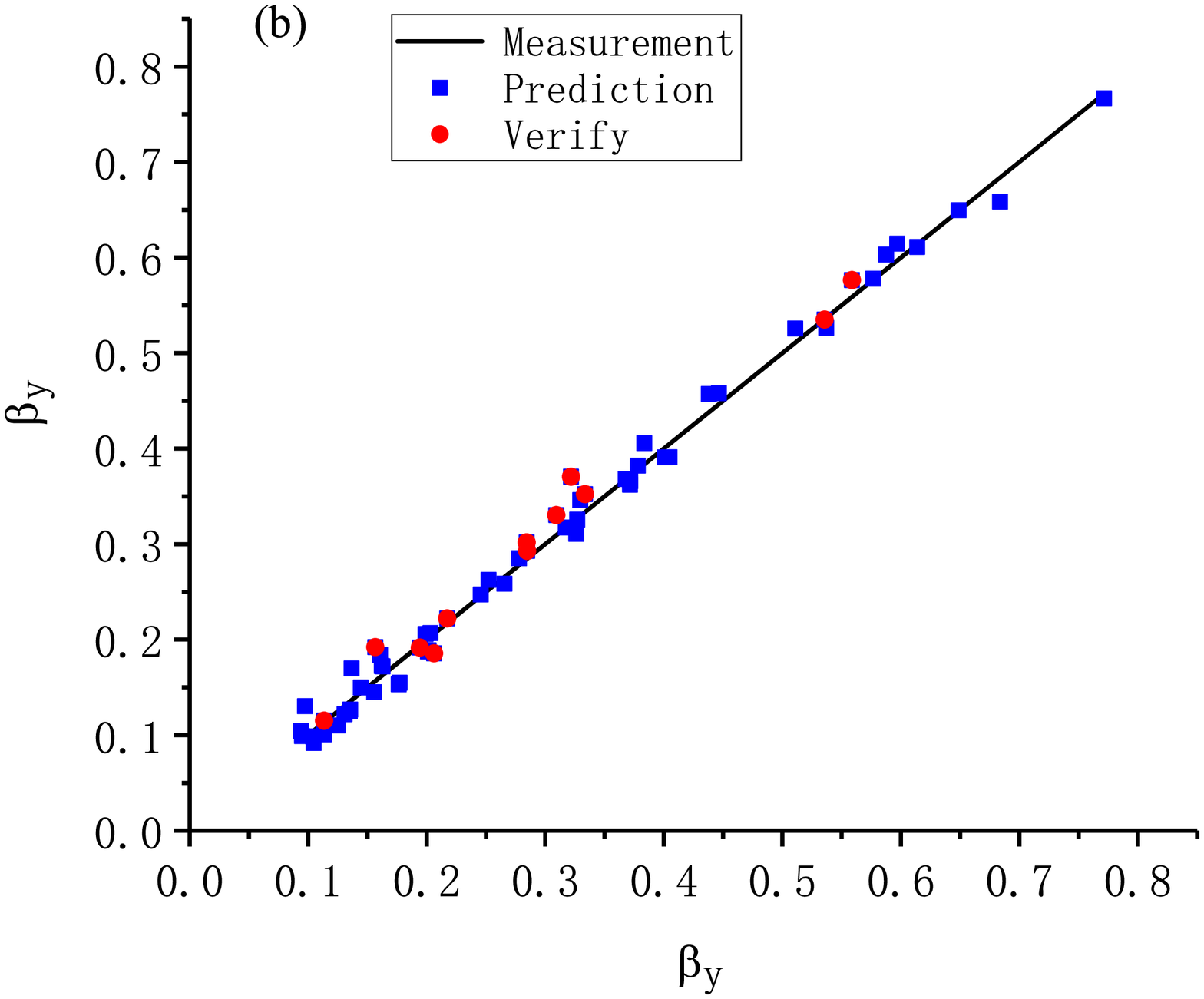}}\vspace{-10pt}
	\caption{The cross-validations of the output neuron (a) $\beta_x$ and (b) $\beta_y$ using the experimental measurements as the training set. The diagonal line represents experimental data, while rectangular (blue) and circle (red) correspond to predication by the model and further test of the model with the measurement in the last module of each block in experimental precedure. }
	\label{verify}
\end{figure}

The loss function is checked in the neural network module, verifying the robustness of the model. Figure ~\ref{loss} is an exemplary. It shows the variation of loss function with iteration number during the training of the model for $\beta_y$. Black curve represents the loss function between measurement and the prediction from the model, while the red curve represents the loss function between verification measurement and the prediction from the model. As is shown, the loss function drops logarithmically with the iteration number. As is seen in the Figure ~\ref{loss} both of them drop to $10^{-3}$ order of magnitude after roughly 700 iterations. This is double-confirmed in Figure ~\ref{verify}\cite{wang2020spcanet}. The measured $beta$ used for tranining, the predicted $beta$ by the trained model using neural network, and the measured $beta$ based on trained model for its verification are all plotted to the measured $beta$ used for tranining. For both $beta_x$ and $beta_y$, the predicted value (blue rectangle), and the verify data (red circle) are in close vicinity with the measurement (black line), proving the accuracy and robustness of the model. 

\begin{figure}
	\centering
	\subfigure{\includegraphics[width=3.3in]{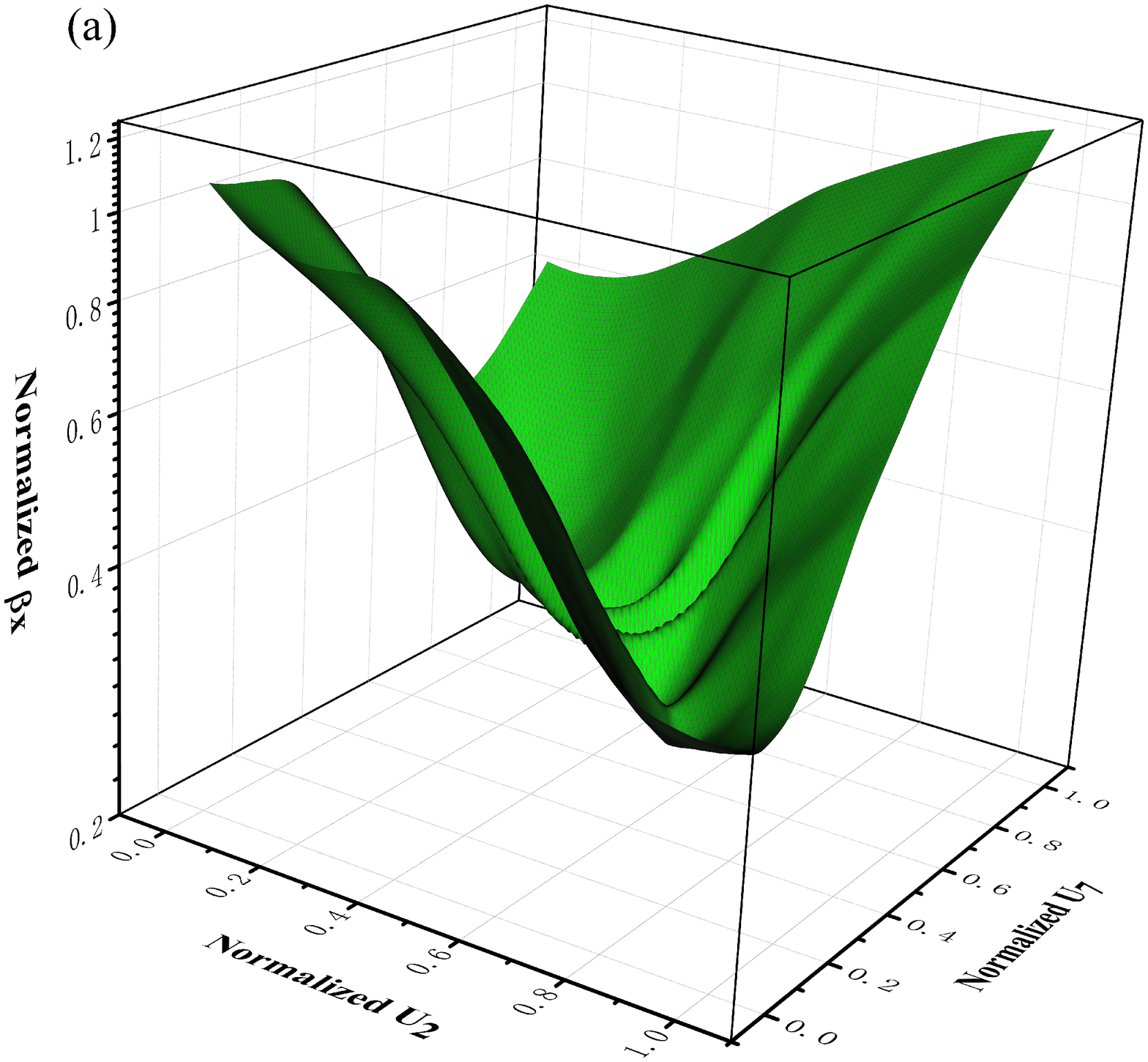}}\vspace{-10pt}
	\subfigure{\includegraphics[width=3.3in]{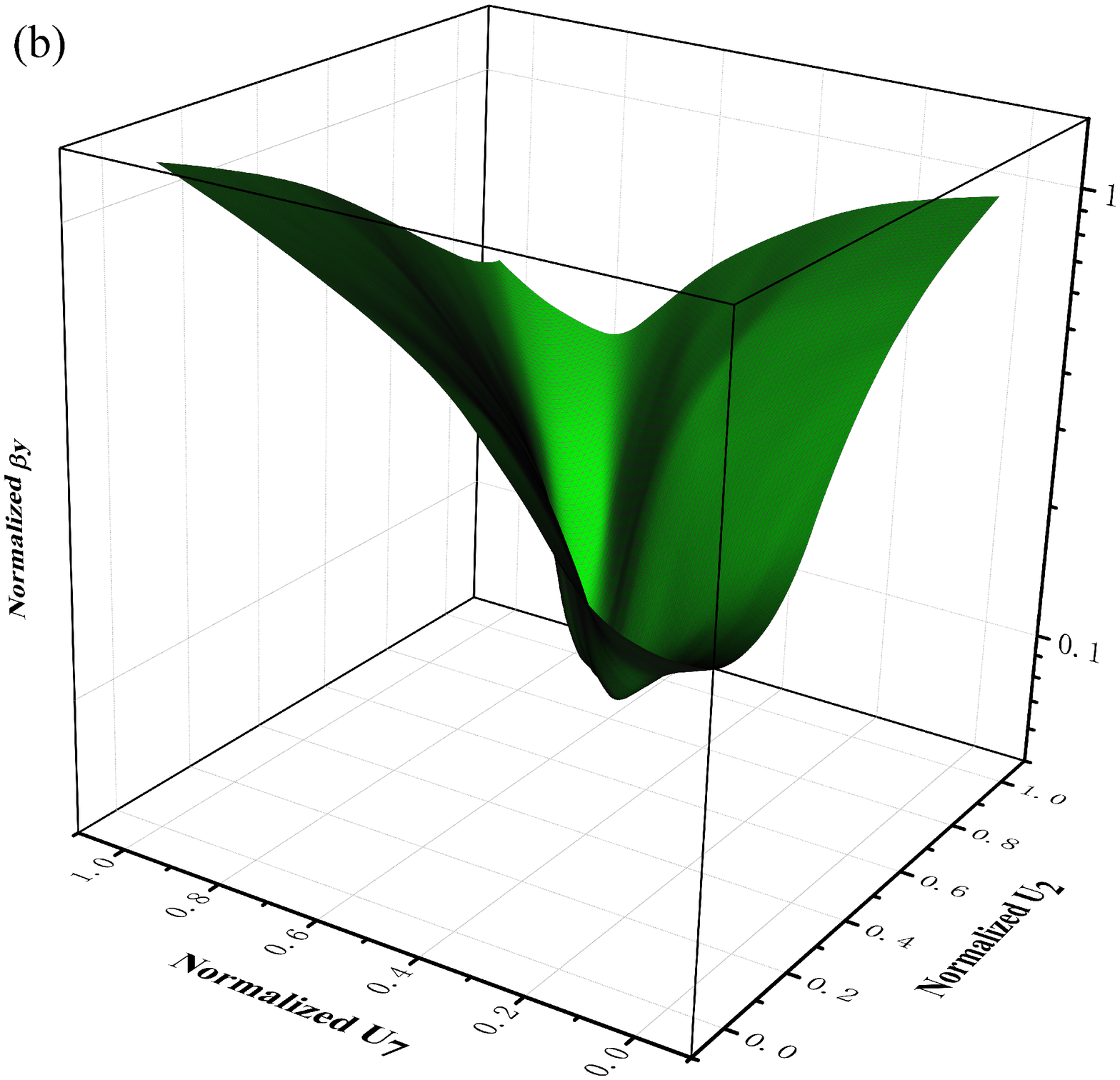}}\vspace{-10pt}
	\caption{The three-dimensional view of the learning model for (a)$\beta_x$ and (b)$\beta_y$, respectively. }
	\label{beta_3d}
\end{figure}

 The results output from the trained model can be visualized by Figure ~\ref{beta_3d}, which shows the three-dimensional view for $\beta_x$ and $\beta_y$ as the function of normalized $U_2$ and $U_7$, respectively. It can be noted that in the investigated parameter space of $U_2$ and $U_7$, both $\beta_x$ and $\beta_y$ has a valley shape, indicating minimum $\beta_x$ and $\beta_y$ exist and have been searched out for proper combination of $U_2$ and $U_7$ by the trained model. 

\begin{figure}
	\centering
	\subfigure{\includegraphics[width=3.3in]{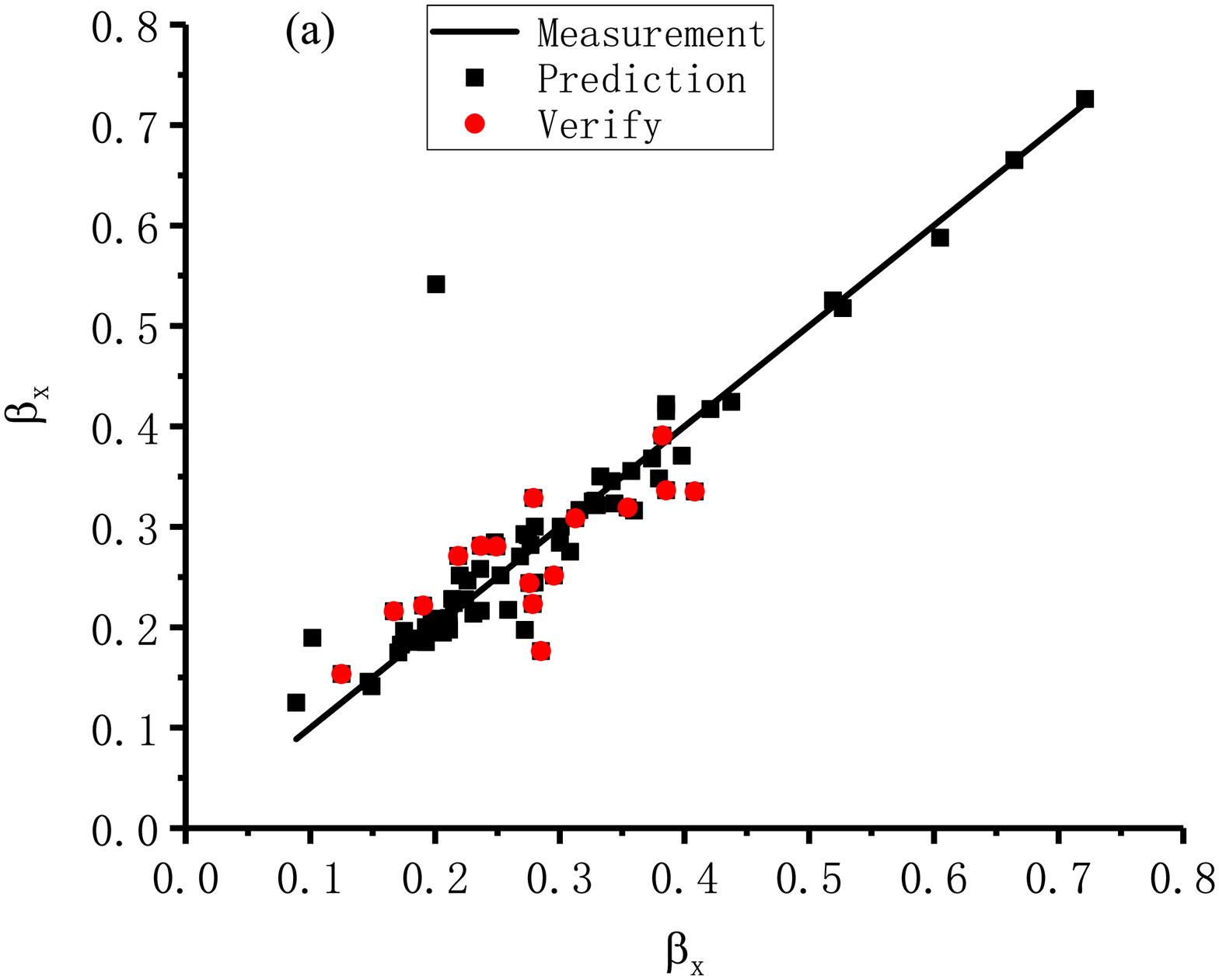}}\vspace{-10pt}
	\subfigure{\includegraphics[width=3.3in]{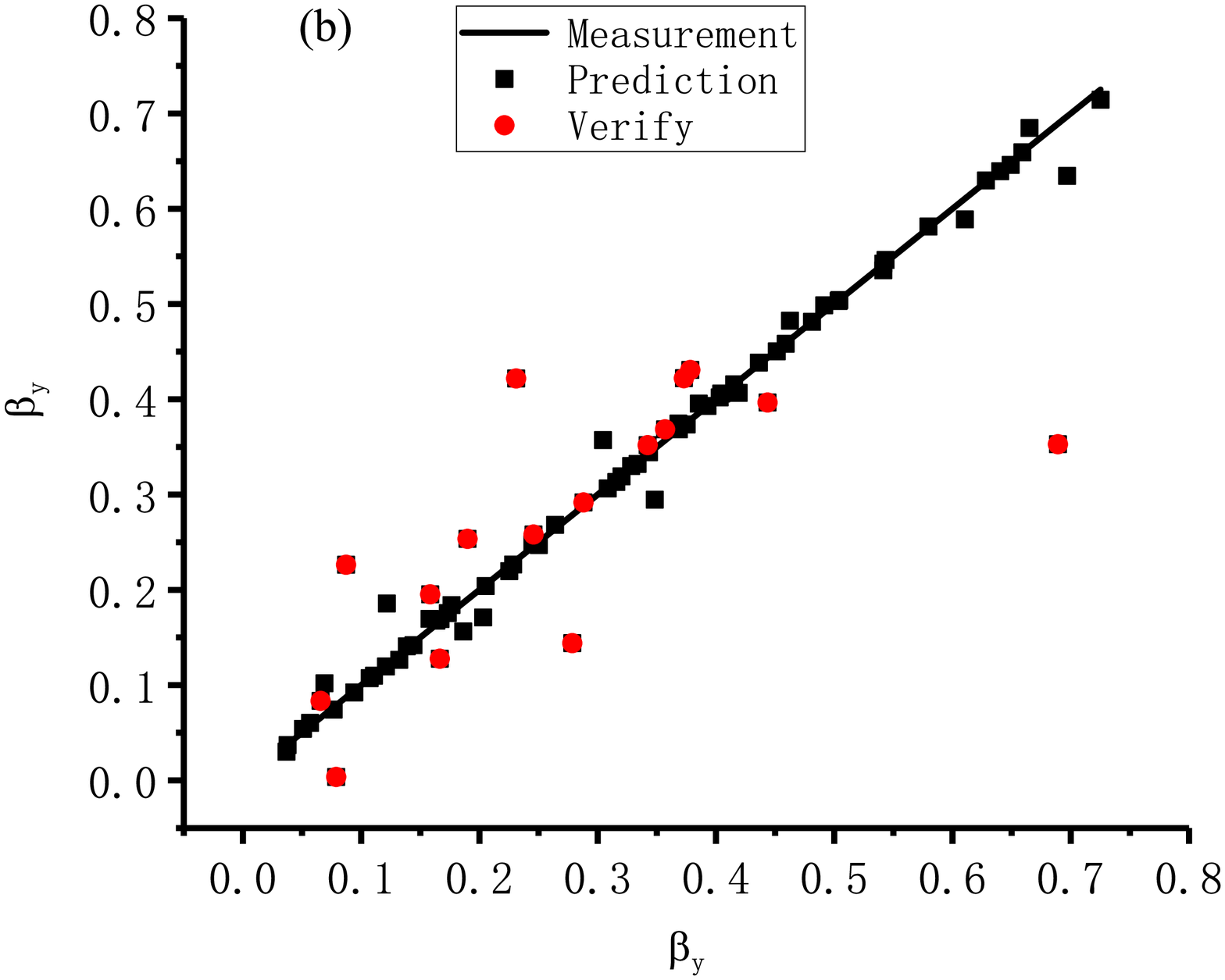}}\vspace{-10pt}
	\caption{The cross-validations of the output neuron (a) $\beta_x$ and (b) $\beta_y$ with three voltage inputs $(U_2, U_3, U_7)$. The diagonal line represents experimental data, while rectangular (blue) and circle (red) correspond to predication by the model and further test of the model with the measurement in the last module of each block in experimental precedure. }
	\label{beta_3_2}
\end{figure}

From these measurements, we performed such proof-of-principle tests using two voltage inputs and two outputs $(\beta_x,\beta_y)$ for the training using neural network. We further extended such tests with three voltage inputs $(U_2, U_3, U_7)$ and four voltage inputs $(U_2, U_3, U_7,U_8)$. With these inputs, we performed similar training to the model using neural network, and verified accuracy of the trained model. As shown in Figure ~\ref{beta_3_2}, the predicted value (blue rectangle), and the verify data (red circle) agree well with the measurement (black line), demonstrating the accuracy of the trained model. 

\begin{figure}[htbp]
	\centering
	\subfigure{\includegraphics[width=3.3in]{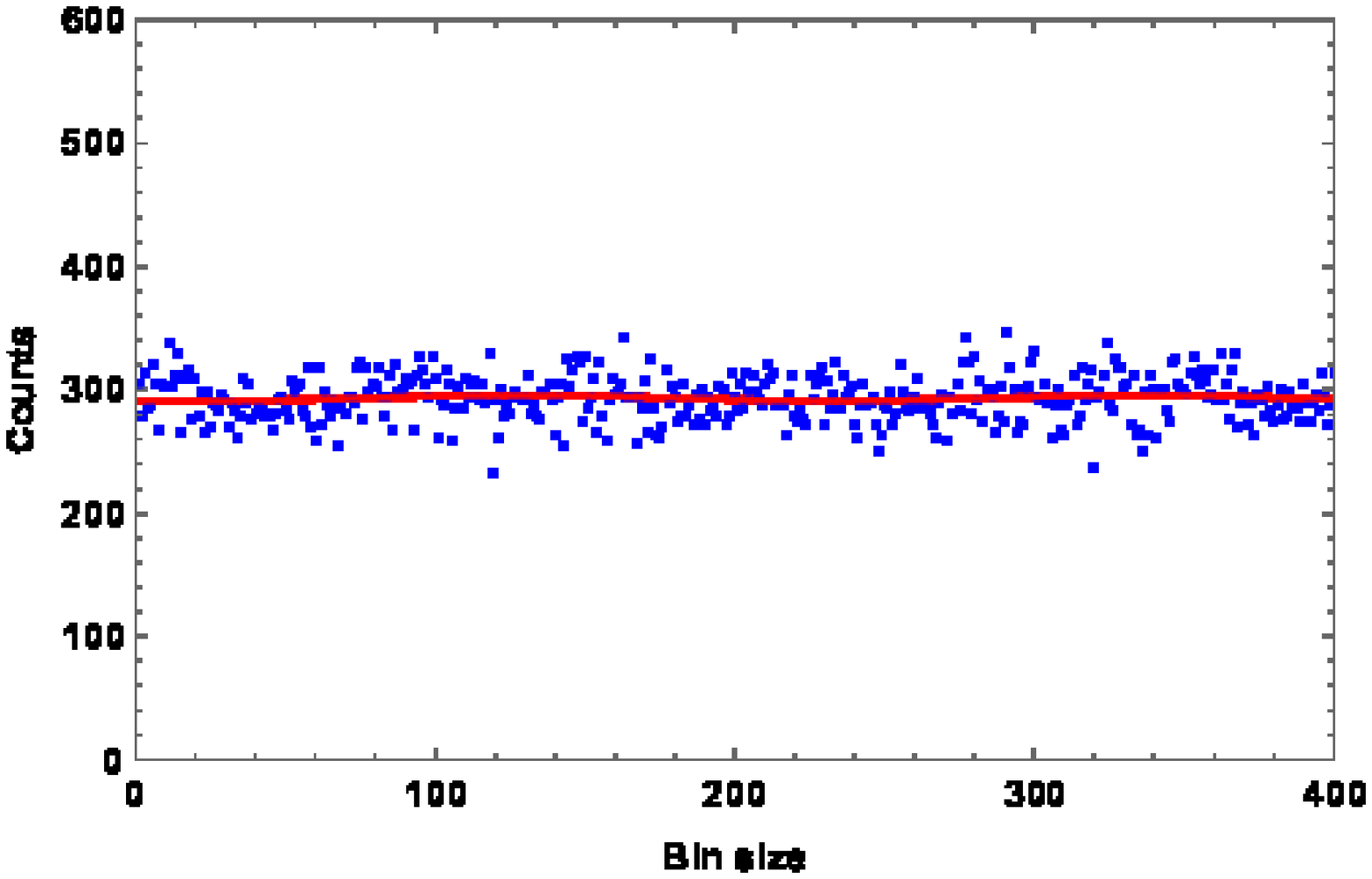}}\vspace{-10pt}
	\subfigure{\includegraphics[width=3.3in]{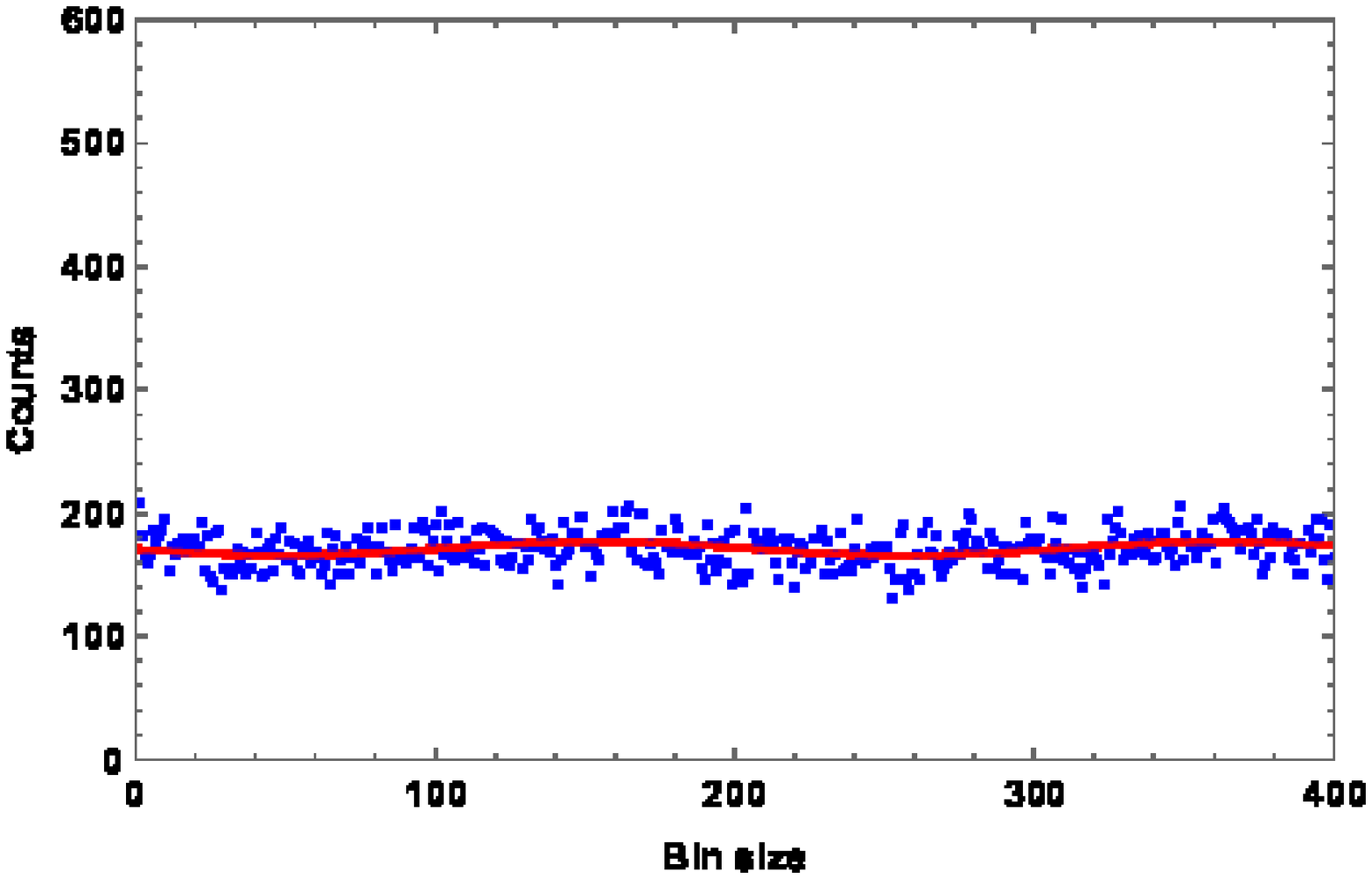}}\vspace{-10pt}
	\caption{The obtained best minimization of micromotion with (a) $\beta_{x, min}$ and (b)$\beta_{y, min}$ during the photon-correlation measurement with our machine-learning approach. }
	\label{index_measurement}
\end{figure}

For four voltage inputs, we found the global minimum of $beta_x$ and $beta_y$ for a specific combination of $(U_2, U_3, U_7,U_8)$. The generated photon-correlation spectrum is shown in Figure ~\ref{index_measurement}. The corresponding $\beta_x$ and $\beta_y$ are inferred to $\beta_x=0.0060\pm0.0026$ and $\beta_y=0.0308\pm0.0028$ by fitting the photon-correlation spectrum in Figure ~\ref{index_measurement} to the Equation \ref{Fitting}. Such order of magnitude show an excellent minimization and is comparable to \cite{1998JAP, 2015JAP}, clearly demonstrating the efficacy of machine learning in this application. In addition, we reached the same results with decision tree, another widely used algorithm in machine learning.

\section*{Conclusion}
we have demonstrated a high-level of micromotion suppression by neural-network based machine learning, using RF-photon correlation technique. With our approach, both $\beta_{x}$ and $\beta_{y}$ have been reduced to $10^{-3}$ order of magnitude, reaching state-of-the-art sensitivity and comparable to the those used for atomic clocks \cite{1998JAP, 2015JAP} and quantum metrology. $\beta_{z}$ is estimated to within $10^{-2}$ according to its relation with $\beta_{y}$, as discussed in the section of theoretical scheme. We also envision that this approach can be generally applied to other similar scenarios in atomic, molecular and optical physics, quantum information processing, and quantum computation. 

\section*{Acknowledgement}
Yang Liu acknowledge the financial support from National Natural Science Foundation of China(NSFC) under Grant No. 11974434, Fundamental Research Funds for the Central Universities of Education of China under Grant No. 191gpy276, Natural Science Foundation of Guangdong Province under Grant 2020A1515011159. Le Luo received supports from NSFC under Grant No.11774436, Guangdong Province Youth Talent Program under Grant No.2017GC010656, Sun Yat-sen University Core Technology Development Fund, and the Key-Area Research and Development Program of GuangDong Province under Grant No.2019B030330001.

\bibliography{ML}

\end{document}